\documentclass[10pt,letterpaper,twocolumn]{article}

\usepackage[top=2.1cm,left=2.0cm,right=2.0cm,footskip=2.0cm]{geometry}
\usepackage[utf8]{inputenc}
\usepackage{cite}
\usepackage{caption}
\usepackage{microtype}
\usepackage{algpseudocode}
\usepackage{algorithm}
\usepackage{graphicx}
\usepackage{url}

\begin{document}

\twocolumn[
    \begin{flushleft}
        {\Large
        \textbf\newline{
            Periortree: An Extention of R-Tree for Periodic Boundary Conditions
            }
        }
        \newline
        \\
        Toru Niina \textsuperscript{1}
        \\
        \bigskip
        \bf{1.} Department of Biophysics, Graduate School of Science,
                Kyoto University, Kyoto 606-8502, Japan
        \\
        \bigskip
        * niina@theory.biophys.kyoto-u.ac.jp
    \end{flushleft}

    \section*{Abstract}
    Searching spatial data is an important operation for scientific simulations
    which are performed mostly with periodic boundary conditions.
    An R-Tree is a well known tree data structure used to contain spatial
    objects and it is capable of answering to spatial searching queries in an
    efficient way.
    In this paper, a novel method to construct an R-Tree considering periodic
    boundary conditions is presented.
    Unlike existing methods, the proposed method works without any kind of
    extra objects or queries.
    Moreover, because this method reduces the volume of bounding box for each
    node under the periodic boundary conditions,
    it is expected to increase the overall efficiency.
    While the extension of an R-Tree is presented in this work, this method is
    not only applicable to an R-Tree but also to other data structures that use
    axis-aligned bounding boxes with periodic boundary conditions.
    The implementation is available on GitHub.
    \bigskip
]

\section*{Introduction}

Computational simulations are essential tools for scientific research, such as
studying behaviors of complex biochemical models\cite{Takahashi2004}.
To perform such large scale simulations, both huge amount of computational
resources and efficient simulation softwares are required.

In most cases, searching for objects that satisfy some geometrical conditions
is one of the most time-consuming operation in a simulation. Generally, an
efficient algorithm that search objects drastically accelerates not only
the whole simulation processes, but also the data analysis of simulation results.
Therefore, a method that efficiently processes spatial search accelerates the
whole process of simulation in scientific research.

An R-Tree is a widely used data structure representing bounding volume
hierarchies (BVH) by using an axis-aligned bounding box (AABB) for all its
entries~\cite{Guttman1984}.
It is capable of containing both sizeless and finite sized objects such as
points, segments, rectangles, spheres, etc.
In order to improve the efficiency, many variants of the R-Tree algorithm have
been proposed~\cite{Greene1989, Beckmann1990, Leuteneggert1997, Mitsuhashi2016}.

\begin{figure}[hbt]
    \includegraphics[width=8.4cm]{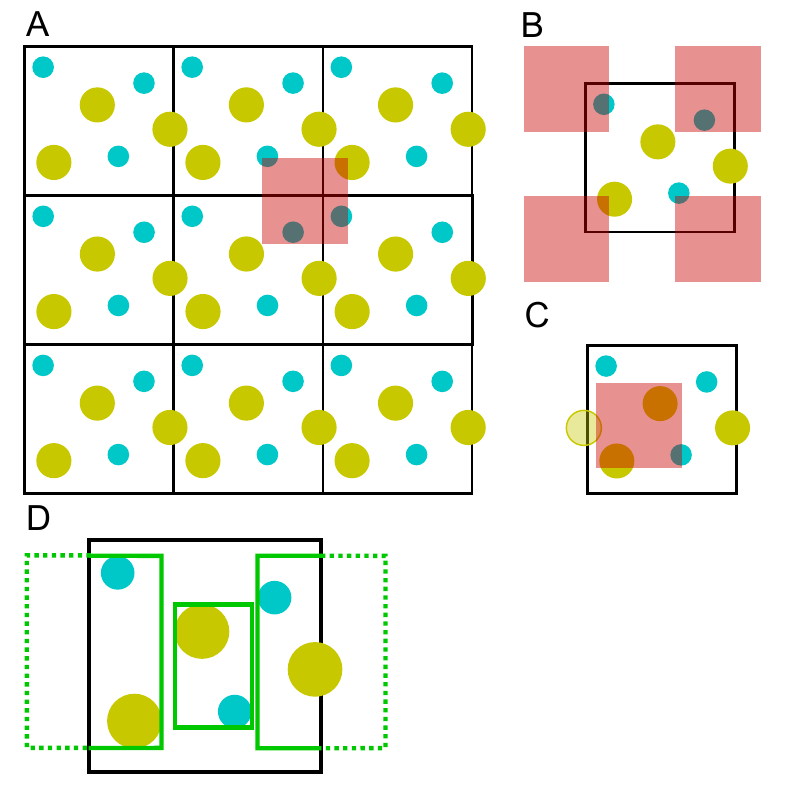}
    \caption{Methods to consider PBCs with R-Tree.
    (\textbf{A})
    By copying the unit cell, the standard R-Tree can manage objects that are
    associated with periodic images.
    (\textbf{B})
    By copying query, R-Tree can find objects that are beyond boundary.
    (\textbf{C})
    With the method described in \textbf{B}, finite sized objects are
    overlooked when queries that is inside of the boundary are not copied.
    (\textbf{D})
    The method proposed in this paper. Forming AABBs according to
    the periodicity, R-Tree can organize objects under periodic boundary
    conditions.}
    \label{fig-method-rtree-pbc}
\end{figure}

In order to use an R-Tree with periodic boundary conditions (PBCs), currently,
there exists two methods in the literature
(Figure~\ref{fig-method-rtree-pbc})~\cite{Mitsuhashi2016}.
The first method is described in Figure~\ref{fig-method-rtree-pbc}A.
It enables an R-Tree to contain and search objects considering PBCs
by storing the periodic images surrounding of the primary unit cell.
However, compared to the standard R-Tree, the memory consumption is increased
$3^D$ times (where $D$ reprecents a dimension of a system).
On the contrary, the other method (Figure~\ref{fig-method-rtree-pbc}B)
does not store any periodic image, but it replicates the query based on the
PBCs by transposing it periodically when the query sticks out of the boundary.
Although it works fine with points, it has a limitation when working with
finite-sized objects because if the query is inside of the boundary and the
objects extend beyond the unit cell, some of them are overlooked
(Figure~\ref{fig-method-rtree-pbc}C).
One ad-hoc solution to overcome this limitation is changing the criteria for
replicating a query based on the size of elements that are contained.
However,  this may increase the frequency in which queries are copied and
decrease the efficiency.

Here a novel method to construct and use an R-Tree with PBCs is proposed.
The main idea is to consider the PBCs when either entries are inserted or
objects are searched.
By expanding an AABB and detecting intersection between AABBs under PBCs,
an R-Tree overcomes the limitation of containing finite-sized object under PBCs
(Figure~\ref{fig-method-rtree-pbc}D). Intersection and inclusion detection can be
implemented by introducing periodic transposition in the existing algorithms
with the appropriate representation of an AABB.
Comparing to the other methods, this method does not require to copy objects or queries.
Moreover, using the information of the PBCs, it has more chance
to reduce the volume of AABBs of each node. Since the spatial
searching of an R-Tree is strongly affected by the volme of each node,
this feature might increase the efficiency.

The codes that are used in this paper are available on GitHub
(\url{http://github.com/ToruNiina/periortree}).

\section*{Methods}

Here some operations to modify and handle rectangles under PBCs are introduced.
To construct a tree structure, Guttman's original quadratic algorithm is
employed~\cite{Guttman1984} without any modification except for expansion of
an AABB and intersection detection.
The fact that this method is independent from the core algorithm to construct
an R-Tree means that it can be applied to most of the R-Tree variants and to
other spatial indexing methods that are based on BVH and AABBs.

Since the rectangles used in an R-Tree are axis-aligned, operation in each
dimension are independent from each other.
It means that it is sufficient to show an operation for one dimension.
It should be noted that AABBs only intersect if they overlap in all the axes.

\paragraph{Representation of PBCs.}
Here, only cuboids are considered as the unit cell of the PBCs.
It is assumed that all the coordinates of the objects are restricted to be
inside of the unit cell.
There are well-known algorithm to restrict positions to be inside of the unit
cell and to find the minimum distance between objects acoording to the PBCs.
In this paper, these algorithms are called as RestrictPosition and
RestrictVector, respectively.

\begin{figure}[thb]
    \includegraphics[width=8.4cm, bb=2 6 226 165]{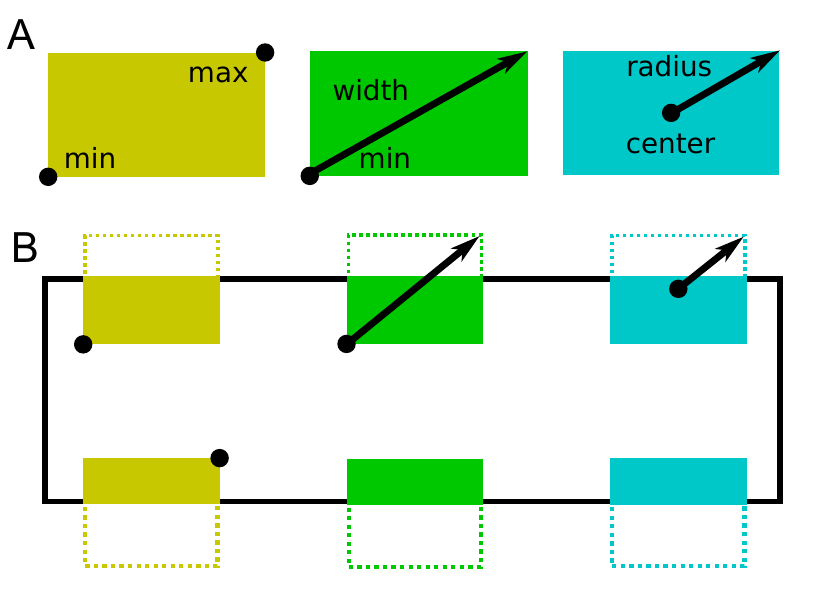}
    \caption{
    (\textbf{A})
    Three popular representations of an AABB. The first one is min-max,
    the second one is min-widths, and the third one is center-radius representation.
    (\textbf{B})
    Three representations in a system with PBCs.
    The AABBs are colored in the same way as \textbf{A}.
    }
    \label{fig-rectangle-rep}
\end{figure}

\paragraph{Representation of AABBs.}
There are three common representations for AABBs
(Figure~\ref{fig-rectangle-rep})~\cite{real-time-collision-detection}.
The first one is by the minimum and maximum coordinate values in each dimension.
The second one is by the minimum point and the diameter from the corner point.
The third one is by the center point and a half of its diameter.
Here, these are called as min-max, min-widths, and center-radius representation,
respectively.

Since the proposed method uses the center point of an AABB, the natural choice
is to use center-radius representation even though the center point can be
calculated by using the other representation.

\paragraph{Expanding AABBs under PBCs.}
Expanding an AABB in order to contain a new object is one of the most
important operations in an R-Tree algorithm.
Because the total area covered by the AABB of each node affects the efficiency
of spatial searching, it is needed to find a way to make it as small as possible.

The algorithm to expand an AABB needed to contain object AABB under PBCs is
shown in Algorithm~\ref{expand_aabb_aabb}.
In this case, the size of an AABB that contains two rectangles is determined
by their radius and the minimum distance between their center points.
In this paper, this is accomplished by using RestrictVector. On the other hand,
when the contents are points, it is enough to find the minimum distance between
the center of an AABB and the point(Algorithm~\ref{expand_aabb_point}).

\begin{algorithm}[tb]
    \caption{expand AABB to contain another AABB}
    \label{expand_aabb_aabb}
    \begin{algorithmic}
        \State $R1 \gets$ AABB to be expanded
        \State $R2 \gets$ rectangle to be contained
        \State $B  \gets$ boundary
        \Function{ExpandAABB}{$R1, R2, B$}
            \State $dc \gets R2.center - R1.center$
            \State $dc \gets$ \Call{RestrictVector}{dc, B}

            \State $l1 \gets R1.center - R1.radius$
            \State $u1 \gets R1.center + R1.radius$
            \State $l2 \gets (R1.center + dc) - R2.radius$
            \State $u2 \gets (R1.center + dc) + R2.radius$

            \State $L  \gets \Call{min}{l1, l2}$
            \State $U  \gets \Call{max}{u1, u2}$
            \State $C  \gets (L + U) / 2$
            \State $HW \gets (U - L) / 2$
            \State $C  \gets$ \Call{RestrictPosition}{C, B}

            \State \Return $\{C, HW\}$
        \EndFunction
     \end{algorithmic}
\end{algorithm}

\begin{algorithm}[tb]
    \caption{expand AABB to contain a point}
    \label{expand_aabb_point}
    \begin{algorithmic}
        \State $R \gets$ AABB to be expanded
        \State $P \gets$ point to be contained
        \State $B \gets$ boundary
        \Function{ExpandAABB}{$R, P, B$}
            \State $dc \gets P - R.center$
            \State $dc \gets$ \Call{RestrictVector}{dc, B}

            \State $l \gets R.center - R.radius$
            \State $u \gets R.center + R.radius$

            \State $L  \gets \Call{min}{l, P}$
            \State $U  \gets \Call{max}{u, P}$
            \State $C  \gets (L + U) / 2$
            \State $HW \gets (U - L) / 2$

            \State $C \gets$ \Call{RestrictPosition}{C, B}
            \State \Return $\{C, HW\}$
        \EndFunction
     \end{algorithmic}
\end{algorithm}

\paragraph{Finding an object under PBCs.}

To find an object in an R-Tree, it is needed to check if two AABBs intersect
each other.
This can be determined by calculating the minimum distance between the centers of
the rectangles(Algorithm~\ref{aabb_intersects_aabb}).
The inclusion can be detected in the same way (Algorithm~\ref{aabb_within_aabb}).
For points, the intersection and inclusion means almost same thing because points
have no size. Here, the algorithm that checks if a point is inside of an AABB
is shown in Algorithm~\ref{point_within_aabb}.

\begin{algorithm}[tb]
    \caption{Check if two AABB intersect each other}
    \label{aabb_intersects_aabb}
    \begin{algorithmic}
        \State $R1 \gets$ rectangle
        \State $R2 \gets$ rectangle that might intersect to R1
        \State $B  \gets$ boundary
        \Function{IntersectsAABB}{$R1, R2, B$}
            \State $dc \gets R1.center - R2.center$
            \State $dc \gets$ \Call{RestrictVector}{dc, B}
            \State \Return \Call{abs}{dc} $\leq (R1.radius + R2.radius)$
        \EndFunction
     \end{algorithmic}
\end{algorithm}

\begin{algorithm}[tb]
    \caption{Check if an AABB is inside of another AABB}
    \label{aabb_within_aabb}
    \begin{algorithmic}
        \State $R1 \gets$ rectangle
        \State $R2 \gets$ rectangle that might be inside of R1
        \State $B  \gets$ boundary
        \Function{IsAABBInsideOfAABB}{$R1, R2, B$}
            \State $dc \gets R1.center - R2.center$
            \State $dc \gets$ \Call{RestrictVector}{dc, B}

            \State \Return \Call{abs}{dc} $\leq (R1.radius - R2.radius)$
        \EndFunction
     \end{algorithmic}
\end{algorithm}

\begin{algorithm}[tb]
    \caption{Check if a point is inside of an AABB}
    \label{point_within_aabb}
    \begin{algorithmic}
        \State $R \gets$ rectangle
        \State $P \gets$ point that might be inside of R
        \State $B \gets$ boundary
        \Function{IsPointInsideOfAABB}{$R, P, B$}
            \State $dc \gets R.center - P$
            \State $dc \gets$ \Call{RestrictVector}{dc, B}
            \State \Return \Call{abs}{dc} $\leq R.radius$
        \EndFunction
     \end{algorithmic}
\end{algorithm}

\section*{Results}

\begin{figure}[tb]
    \includegraphics[width=8.4cm, bb=3 6 237 282]{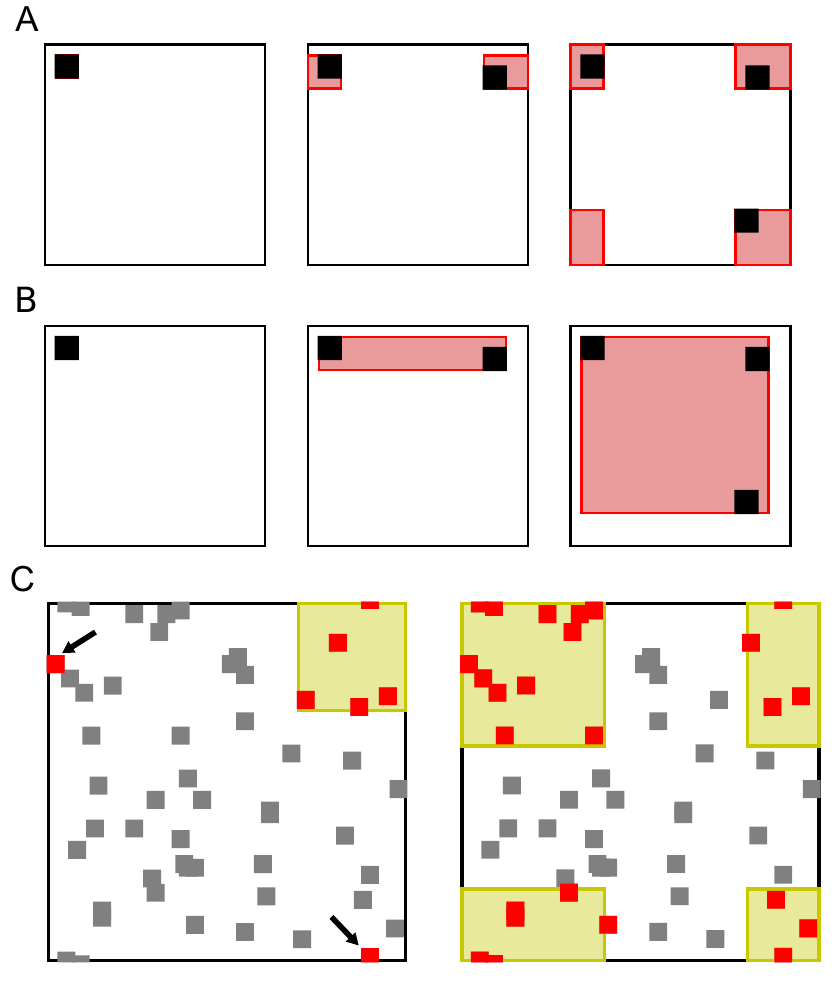}
    \caption{
    (\textbf{A})
    Expansion of the AABB(red) by proposed method.
    It can be seen that the AABB is expanded beyond the boundary.
    (\textbf{B})
    Expansion of the AABB by existing method.
    The AABB wraps almost all the area of the system
    because the AABB is expanded independently from PBCs.
    (\textbf{C})
    The result of querying objects that intersect with the yellow rectangle.
    The objects contained in the R-Tree are shown as small square boxes
    and the ones intersecting the query box are colored in red.
    }
    \label{fig-result}
\end{figure}

Figure~\ref{fig-result}A shows the expansion of an AABB to contain objects by
using proposed method and Figure~\ref{fig-result}B shows that of existing method.
There, the objects are colored in black, and the AABB is colored in red.
It should be noted that, because it sticks out of the boundaries,
the periodic images of the AABB are shown separately.

Comparing Figure~\ref{fig-result}A and B, it is shown that the proposed method
successfully finds least expansion under PBCs.
Because the objects are distributed throughout the system,
if the AABB was expanded independently from the PBCs,
the AABB would wrap most of the area of the system(Figure~\ref{fig-result}B).

In Figure~\ref{fig-result}C the results of a query are shown.
It successfully finds objects that are located beyond the boundaries(indicted by arrows).
As in Figure~\ref{fig-result}A, the periodic images of the query are
shown separately.

\section*{Conclusion}

In this paper, a novel method for applying PBCs to R-Trees was presented.
This method considers PBCs while inserting new objects and finding objects.
Therefore, no extra objects or queries are needed to consider PBCs.
Moreover, allowing AABBs to extend beyond the boundary, it has more chance to
reduce the volume of AABBs based on the PBCs.

As described before, since this method modifies only the algorithms to expand an
AABB and to detect intersection between AABBs, it is essentially applicable to
other R-Tree variants and to any spatial indexing method that works with AABBs.

Because the application of spatial indexing generally increase the performance
of spatiotemporal simulations, the proposed method might be useful for
scientific simulations under PBCs.

\bibliographystyle{unsrt}
\bibliography{library}{}

\end{document}